\documentclass[11pt,a4paper]{article}
\newcommand{\be}{\begin{equation}}
\newcommand{\ee}{\end{equation}}

\def\mt{{\mbox{\tiny{(1)}}}}
\def\my{{\mbox{\tiny{(2)}}}}
\def\mp{{\mbox{\tiny{(3)}}}}
\def\mq{{\mbox{\tiny{(4)}}}}
\def\mf{{\mbox{\tiny{(5)}}}}
\def\ms{{\mbox{\tiny{(6)}}}}
\newcommand{\Gammabol}{\Gamma{}}
\newcommand{\Rbol}{R{}}

\newcommand{\onehalf}{{\textstyle{\frac{1}{2}}}}

\renewcommand{\thefootnote}{\fnsymbol{footnote}}

\usepackage{revsymb}
\usepackage{bm}
\usepackage{amssymb}
\usepackage{amsmath}

\usepackage{color}
\RequirePackage[numbers,sort&compress]{natbib}

\begin{document}

\renewcommand{\thefootnote}{\fnsymbol{footnote}}
\noindent
{\Large \bf Exploring Higher-Order Gravitational Waves}
\vskip 0.7cm
\noindent
{\bf H. Arcos,$^{\rm a}$ M. Kr\v{s}\v{s}\'ak$^{\rm b}$ and J. G. Pereira$^{\rm b}$}
\vskip 0.2cm \noindent
$^{\rm a}${\it Universidad Tecnol\'ogica de Pereira\\
A.A. 97, La Julita, Pereira, Colombia}
\vskip 0.2cm \noindent
$^{\rm b}${\it Instituto de F\'{\i}sica Te\'orica,
Universidade Estadual Paulista \\
Caixa Postal 70532-2, 01156-970 S\~ao Paulo, SP, Brazil}

\vskip 0.8cm

\begin{quote}
{\bf Abstract.}~{\footnotesize In addition to the usual linear gravitational waves in transverse-traceless coordinates, higher-order gravitational field equations, as well as their corresponding solutions, are explicitly obtained. It is found that higher-order waves do not represent corrections to the first-order wave. In contrast, all higher than second-order solutions do represent corrections to the second-order wave, a property that makes the first-order gravitational wave to stand apart from higher-order waves. Furthermore, although the first-order solution is transverse and traceless, all higher-order solutions are not. As a consequence, the whole solution is neither transverse nor traceless, a result that could eventually have important consequences for quantum gravity, and in particular for the definition of graviton itself. Some additional properties and features of these higher-order gravitational waves are explored and discussed.

}

\end{quote}

\section{Introduction}

Although there seems to be a general agreement that the transport of energy by gravitational waves is a non-linear phenomenon \cite{transport}, due to the tiny amount of energy transported, it is usual to assume that the linear approximation is able to accurately describe the propagation of gravitational waves \cite{maggiore}. For this reason, when studying gravitational waves far away from the sources---that is, in the so-called wave zone---only the linear approximation is considered. Denoting by $h^{\,\mu}_{\mt \nu}$ the first-order term of the post-Minkowskian expansion, and choosing the class of harmonic coordinates, in which $h^{\,\mu}_{\mt \nu}$ satisfies the condition
\be
\partial_\mu h^{\,\mu}_{\mt \nu} - {\textstyle{\frac{1}{2}}} \partial_\nu h_\mt = 0,
\label{1oHC}
\ee
with $h_\mt = h^{\,\mu}_{\mt \mu}$, the first-order vacuum field equation assumes the form
\be
\Box h^{\,\mu}_{\mt \nu} = 0.
\ee
Gravitational waves are usually interpreted as a solution to this set of equations.

On the other hand, higher-order gravitational waves are in general studied through the use of an induction process, which amounts to consider non-linear iteration of the field equations. Such approach yields, for any $n \geq 2$, the set of higher-order equations \cite{living}
\be
\partial_\mu h^{\,\mu}_{(n) \nu} - {\textstyle{\frac{1}{2}}} \partial_\nu h_{(n)} = 0
\label{hoHC}
\ee
\be
\Box h^{\,\mu}_{(n) \nu} = \Lambda^{\,\mu}_{(n) \nu}[h_\mt, h_\my, \dots, h_{(n-1)}].
\label{hoFE}
\ee
Even though the general mathematical properties of such system have been extensively explored in the literature \cite{math1}, a thorough study of the explicit form of the field equations for $n \geq 2$, as well as their corresponding solutions, is still lacking. With the purpose of getting a clearer and more precise view of higher-order gravitational waves, we are going in this paper to explicitly obtain a few higher-order field equations along with the corresponding coordinate conditions. Then, by solving these equations, some features and properties of higher-order gravitational waves will be studied and discussed.

\section{Gravitational waves}
\label{main}

\subsection{Perturbative expansion}

Considering that the sources of gravitational waves are at enormous distances from Earth, it is sensible to assume that the amplitude of a gravitational wave when reaching a detector on Earth will be very small. This allows the use of a perturbative analysis in which the gravitational field variable---that is, the metric tensor---is expanded in the form
\be
g_{\mu \nu} = \eta_{\mu \nu} + \varepsilon \, h_{\mt \mu \nu} +
\varepsilon^2 h_{\my \mu \nu} + \cdots
\ee
where $\varepsilon$ is a small parameter introduced to label the successive orders, and $\eta_{\mu \nu}$ is the metric of the background Minkowski spacetime, where the gravitational wave is supposed to propagate. The expansion of the inverse metric is
\be
g^{\mu \nu} = \eta^{\mu \nu} - \varepsilon \, h_\mt^{\,\mu \nu} +
\varepsilon^2 \Big(h_\my^{\mu \nu} + h_\mt^{\,\mu}{}_\alpha h_\mt^{\,\alpha \nu} \Big) 
+ \cdots \; .
\ee
At the first order in the metric perturbation, the Christoffel connection assumes the form
\be
\Gammabol_{{\mbox{\tiny{(1)}}}}^\alpha{}_{\beta \gamma} = 
\onehalf \, \eta^{\alpha \rho} \left(\partial_\beta h_{\mt\rho \gamma} +
\partial_\gamma h_{\mt\rho \beta} - \partial_\rho h_{\mt \beta \gamma} \right)
\label{ChristoConne1}
\ee
whereas the first-order Riemann tensor is
\be
\Rbol_{\mbox{\tiny{(1)}}}^\alpha{}_{\mu \nu \beta} =
\partial_\nu \Gammabol_{\mt}^\alpha{}_{\mu \beta} -
\partial_\beta \Gammabol_{{\mbox{\tiny{(1)}}}}^\alpha{}_{\mu \nu}.
\label{R1}
\ee
At the second order, the Christoffel connection is
\be
\Gammabol^\alpha_{\my \beta \gamma} = 
\onehalf \, \eta^{\alpha \rho} \Big(\partial_\beta h_{\my \rho \gamma} +
\partial_\gamma h_{\my \rho \beta} - \partial_\rho h_{\my \beta \gamma} \Big) -
\onehalf \, h_\mt^{\alpha \rho} \Big(\partial_\beta h_{\mt\rho \gamma} +
\partial_\gamma h_{\mt\rho \beta} - \partial_\rho h_{\mt \beta \gamma} \Big)
\label{ChristoConne2}
\ee
with the second-order Riemann tensor given by
\be
R^{\alpha}_{\my \mu \nu \beta} = \partial_\nu \Gamma^\alpha_{\my \mu \beta} -
\partial_\beta \Gamma^\alpha_{\my \mu \nu} +
\Gamma^\alpha_{\mt \gamma \nu} \Gamma^\gamma_{\mt \mu \beta} -
\Gamma^\alpha_{\mt \gamma \beta} \Gamma^\gamma_{\mt \mu \nu}
\label{R2}
\ee
At the third order, the Christoffel connection is
\begin{eqnarray}
\Gammabol^\alpha_{\mp \beta \gamma} = 
\onehalf \, \eta^{\alpha \rho} \Big(\partial_\beta h_{\mp \rho \gamma} +
\partial_\gamma h_{\mp \rho \beta} - \partial_\rho h_{\mp \beta \gamma} \Big) -
\onehalf \, h_\mt^{\alpha \rho} \Big(\partial_\beta h_{\my\rho \gamma} +
\partial_\gamma h_{\my\rho \beta} - \partial_\rho h_{\my \beta \gamma} \Big) \nonumber \\
+ \onehalf \Big(h_\my^{\alpha \rho} + h_\mt^{\,\alpha}{}_\lambda h_\mt^{\,\lambda \rho} \Big)
\Big(\partial_\beta h_{\mt\rho \gamma} +
\partial_\gamma h_{\mt\rho \beta} - \partial_\rho h_{\mt \beta \gamma} \Big)~~~~~~
\label{ChristoConne3}
\end{eqnarray}
and the third-order Riemann tensor given by
\be
R^{\alpha}_{\mp \mu \nu \beta} = \partial_\nu \Gamma^\alpha_{\mp \mu \beta} -
\partial_\beta \Gamma^\alpha_{\mp \mu \nu} +
\Gamma^\alpha_{\mt \gamma \nu} \Gamma^\gamma_{\my \mu \beta} -
\Gamma^\alpha_{\mt \gamma \beta} \Gamma^\gamma_{\my \mu \nu} +
\Gamma^\alpha_{\my \gamma \nu} \Gamma^\gamma_{\mt \mu \beta} -
\Gamma^\alpha_{\my \gamma \beta} \Gamma^\gamma_{\mt \mu \nu}.
\label{R3}
\ee
And so on, to any higher order.

\subsection{First-order gravitational waves}
\label{TTcs}

In the linear, or first-order approximation, the sourceless gravitational field equation is
\begin{equation}
R_{\mt \mu \nu} = 0
\label{dH1}
\end{equation}
with $R_{\mt \mu \nu}$ the first-order Ricci tensor. In terms of the metric perturbation, it assumes the form
\be
\Box h_{\mt \mu \nu} -
\partial_\mu (\partial_\alpha h_\mt^{\,\alpha}{}_\nu - \onehalf \partial_\nu h_\mt ) -
\partial_\nu (\partial_\alpha h_\mt^{\,\alpha}{}_\mu - \onehalf \partial_\mu h_\mt ) = 0
\label{dH1bis}
\ee
where the notation $h_\mt = h^{\,\mu}_{\mt \mu}$ has been used. 
The class of harmonic coordinates is defined by the condition
\be
g^{\mu \nu} \, \Gamma^\rho{}_{\mu \nu} = 0.
\label{HarmoCoor}
\ee
At the first order it becomes
\be
\partial_\mu h^{\,\mu}_{\mt \nu} - {\textstyle{\frac{1}{2}}} \partial_\nu h_\mt = 0.
\label{hcc}
\ee
In these coordinates, the field equation (\ref{dH1bis}) assumes the form
\be
\Box \, h^{\,\mu}_{\mt \nu} = 0
\label{we1}
\ee
with $\Box$ the flat-spacetime d'Alambertian operator.

A monochromatic plane-wave solution to this equation is
\be
h^{\,\mu}_{\mt \nu} = A^{\,\mu}_{\mt \nu}  \exp[{i k_{\rho} x^\rho}],
\label{pw}
\ee
where $A^{\,\mu}_{\mt \nu}$ is the symmetric polarization tensor. In this case, the wave vector $k^\rho$ is found to satisfy the dispersion relation
\be
k_{\rho}  k_{}^\rho = 0
\label{12}
\ee
and the harmonic coordinate condition (\ref{hcc}) becomes
\be
k_\mu  h^{\,\mu}_{\mt \nu} = {\textstyle{\frac{1}{2}}} k_\nu h_\mt.
\label{13}
\ee
Within the class of harmonic coordinates, it is possible to choose a specific coordinate system. Once this is done, the coordinates become completely specified, and the remaining components of $h^{\,\mu}_{\mt \nu}$ turn out to represent the physical degrees of freedom. A quite convenient choice is the so-called {transverse-traceless} coordinates (or {\em gauge}, in analogy to electromagnetism), in which
\be
h_\mt = 0 \qquad \mbox{and} \qquad
h^{\,\mu}_{\mt \nu} U_{\mbox{\tiny{(0)}}}^\nu = 0
\label{TTA}
\ee
with $U_{\mbox{\tiny{(0)}}}^\nu$ an arbitrary, constant four-velocity. In this coordinate system, condition (\ref{13}) assumes the form
\be
k_\mu  h^{\,\mu}_{\mt \nu} = 0.
\label{13bis}
\ee
Now, although the coordinate system has already been completely specified, we still have the freedom to choose different local frames $e^a = e^a{}_\mu dx^\mu$.
In particular, it is always possible to choose a specific frame in which $U_{\mbox{\tiny{(0)}}}^\nu = \delta_0^\nu$. In this frame, as 
can be seen from the second of the Eqs.~(\ref{TTA}),
\be
h^{\,\mu}_{\mt 0} = 0
\ee
for all $\mu$. Orienting the frame in such a way that the wave travels in the $z$ direction, the wave vector assumes the form
\be
k^\rho = \left({\omega}/{c}, 0, 0, {\omega}/{c} \right),
\label{kzdir}
\ee
and the physical components of the wave are found to be
\be
h^{\,x}_{\mt x} = - h^{\, y}_{\mt y} \qquad \mbox{and} \qquad
h^{\,x}_{\mt y} = h_{\mt y}{}^{x}.
\label{FOmetric}
\ee
Linear waves satisfying these conditions are said to represent a plane gravitational wave in transverse-traceless coordinates.

\subsection{Second-order gravitational waves}

At the second order, the sourceless gravitational field equation acquires the form
\begin{equation}
R_{\my \mu \nu} = 0
\label{dH2}
\end{equation}
with $R_{\my \mu \nu}$ the second-order Ricci tensor. In terms of the components of the metric perturbation, and using for $h_{\mt}^{\mu}{}_{\nu}$ the plane wave solution (\ref{pw}) along with the constraints (\ref{12}) and (\ref{13}), the wave equation (\ref{dH2}) assumes the form
\be
\Box h_{\my \mu \nu} -
\partial_\mu (\partial_\alpha h_\my^{\,\alpha}{}_\nu - \onehalf \partial_\nu h_\my ) -
\partial_\nu (\partial_\alpha h_\my^{\,\alpha}{}_\mu - \onehalf \partial_\mu h_\my ) = -
\textstyle \frac{3}{2} \, \Phi_\my k_\mu k_\nu \exp[{i 2 k_\rho x^\rho}]
\label{dH2bis}
\ee
where $h_\my = h_\my^{\,\alpha}{}_\alpha$ and
\be
\Phi_\my = A^{\,\rho}_{\mt \sigma} A^{\,\sigma}_{\mt \rho}.
\label{fi2}
\ee
On the other hand, at this order the harmonic coordinate condition (\ref{HarmoCoor}) is
\be
\eta^{\mu \nu} \, \Gamma_\my^\lambda{}_{\mu \nu} -
h_\mt^{\mu \nu} \, \Gamma_\mt^\lambda{}_{\mu \nu} = 0.
\ee
Substituting Eqs.~(\ref{ChristoConne1}) and (\ref{ChristoConne2}), it becomes
\be
\partial_\mu h_\my^{\,\mu}{}_\lambda - {\textstyle{\frac{1}{2}}} \partial_\lambda h_\my =
- {\textstyle{\frac{i}{2}}} \Phi_\my k_\lambda \exp[{i 2 k_\rho x^\rho}].
\label{hc2}
\ee
In these coordinates, and using for $h_{\mt}^{\mu}{}_{\nu}$ the plane wave solution (\ref{pw}), as well as the constraints (\ref{12}) and (\ref{13}), the second-order wave equation assumes the form
\begin{equation}
\Box\,  h_{\my}^{\mu}{}_{\nu} = \frac{\Phi_\my}{2} \, k^\mu k_\nu
\exp[i 2 k_\rho x^\rho].
\label{ffee}
\end{equation}
A monochromatic wave solution to this equation is \cite{gw2}
\be
h_{\my}^{\, \mu}{}_{\nu} = \left(A_{\my}^{\,\mu}{}_{\nu} + i B_{\my}^{\,\mu}{}_{\nu} \right)
\exp[i 2 k_\rho x^\rho]
\label{PhysGraWave}
\ee
where
\be
A_{\my}^{\,\mu}{}_{\nu} = \frac{3}{16} \, \Phi_\my \, \delta^\mu_\nu
\qquad \mbox{and} \qquad 
B_{\my}^{\,\mu}{}_{\nu} = - \, \frac{\Phi_\my}{8} \, 
\frac{K_\theta x^\theta}{K_\sigma k^\sigma} \, k^\mu k_\nu
\label{B2}
\ee
are the second-order amplitudes, with $K_\theta$ an arbitrary wave number four-vector.
The amplitude of the first part of the solution satisfies
\be
A_\my^{\,\mu}{}_{\mu} \equiv A_\my = \textstyle{\frac{3}{4}} \, \Phi_\my \qquad \mbox{and} \qquad
k_\mu A_{\my}^{\,\mu}{}_{\nu} = \frac{1}{4} \, k_\nu A_\my
\label{A2vinculo}
\ee
from where we see that, at this order, the solution is no longer traceless. On the other hand, the amplitude of the second part of the solution satisfies
\be
B_\my^{\,\mu}{}_{\mu} \equiv B_\my = 
0 \qquad \mbox{and} \qquad k_\mu B_{\my}^{\,\mu}{}_{\nu} = 0.
\label{ttPart}
\ee
This part of the solution is thus traceless and transverse.

We consider now a laboratory proper frame~---~endowed with a Cartesian coordinate system~---~from which the wave will be observed. If the wave is traveling in the $z$ direction of the Cartesian system, the wave vector is that given by Eq.~(\ref{kzdir}). In this case, the amplitude $A_{\my}^{\mu}{}_{\nu}$ turns out to be
\begin{gather}
\left( A_{\my}^{\,\mu}{}_{\nu} \right) = \frac{3 \Phi_\my}{16} \left(
\begin{matrix}
 1 & 0 & 0 & 0 \\
 0 &1 & 0 & 0 \\
 0 & 0 & 1 & 0 \\
 0 & 0 & 0 & 1
\end{matrix}
\right)
\end{gather}
with each component given by
\be
A^{\;t}_{\my t} = A^{\;x}_{\my x} = A^{\;y}_{\my y} = A^{\;z}_{\my z} \equiv \textstyle{\frac{3}{16}} \, \Phi_\my.
\ee
On the other hand, choosing the arbitrary wave vector $K_\rho$ in such a way that $K_0 = K_1 = K_2 = 0$,\footnote{This choice corresponds to assuming that the wave amplitude $B_{\my}^{\mu}{}_{\nu}$ depends on the distance $z$ from the source, but not on the time.} the amplitude $B_{\my}^{\mu}{}_{\nu}$ turns out to be
\begin{gather}
\left( B_{\my}^{\,\mu}{}_{\nu} \right) = - \, \frac{\Phi_\my z \, \omega}{8 c} \left(
\begin{matrix}
 1 & 0 & 0 & -1 \\
 0 & 0 & 0 & 0 \\
 0 & 0 & 0 & 0 \\
 1 & 0 & 0 & -1
\end{matrix}
\right).
\label{B2z}
\end{gather}
This part of the solution has two physical components, which are given by
\be
B^{0}_{\my 0} = - B^{z}_{\my z} \qquad \mbox{and} \qquad B^{0}_{\my z} = -
B_{\my z}{}^{0}.
\label{PhysCom0}
\ee

\subsection{Third-order gravitational waves}

Following the same steps of the previous section, we consider the third-order sourceless gravitational field equation
\begin{equation}
R_{\mp \mu \nu} = 0
\label{dH3}
\end{equation}
with $R_{\mp \mu \nu}$ the third-order Ricci tensor. In terms of the components of the metric perturbation, and using the solutions already obtained for $h_{\mt}^{\mu}{}_{\nu}$ and  $h_{\my}^{\mu}{}_{\nu}$, the field equation (\ref{dH3}) assumes the form
\be
\Box h_{\mp \mu \nu} -
\partial_\mu (\partial_\alpha h_\mp^{\,\alpha}{}_\nu - \onehalf \partial_\nu h_\mp ) -
\partial_\nu (\partial_\alpha h_\mp^{\,\alpha}{}_\mu - \onehalf \partial_\mu h_\mp ) = -
\, \Phi_\mp k_\mu k_\nu \exp[{ i 3 k_\rho x^\rho}]
\label{dH3bis}
\ee
where $h_\mp = h_\mp^{\,\alpha}{}_\alpha$ and
\be
\Phi_\mp = A^{\,\rho}_{\mt \sigma} A^{\,\sigma}_{\mt \lambda} A^{\,\lambda}_{\mt \rho}.
\label{fi3}
\ee
On the other hand, at this order the harmonic coordinate condition (\ref{HarmoCoor}) is
\be
\eta^{\mu \nu} \, \Gamma_\mp^\lambda{}_{\mu \nu} -
h_\mt^{\mu \nu} \, \Gamma_\my^\lambda{}_{\mu \nu} +
\big(h_\my^{\mu \nu} + h^{\,\mu}_{\mt \rho} h_\mt^{\rho \nu} \big) \Gamma_\mt^\lambda{}_{\mu \nu} = 0.
\ee
Substituting Eqs.~(\ref{ChristoConne1}), (\ref{ChristoConne2}) and (\ref{ChristoConne3}), it becomes
\be
\partial_\mu h_\my^{\,\mu}{}_\lambda - {\textstyle{\frac{1}{2}}} \partial_\lambda h_\my =
{\textstyle{\frac{i}{2}}} \Phi_\mp k_\lambda \exp[{ i 3 k_\rho x^\rho}].
\label{hc3}
\ee
In these coordinates, the third-order wave equation is found to be
\begin{equation}
\Box\,  h_{\mp}^{\mu}{}_{\nu} = - \, \Phi_\mp \, k^\mu k_\nu
\exp[{ i 3 k_\rho x^\rho}].
\label{ffee3}
\end{equation}
A monochromatic wave solution to this equation is
\be
h_{\mp}^{\, \mu}{}_{\nu} = \left(A_{\mp}^{\,\mu}{}_{\nu} + i B_{\mp}^{\,\mu}{}_{\nu} \right)
\exp[{ i 3 k_\rho x^\rho}]
\label{PhysGraWave3}
\ee
where
\be
A_{\mp}^{\,\mu}{}_{\nu} = \frac{2}{9} \, \Phi_\mp \, \delta^\mu_\nu
\qquad \mbox{and} \qquad 
B_{\mp}^{\,\mu}{}_{\nu} = \frac{\Phi_\mp}{6} \, 
\frac{K_\theta x^\theta}{K_\sigma k^\sigma} \, k^\mu k_\nu
\label{B3}
\ee
are the third-order amplitudes, with $K_\theta$ an arbitrary wave number four-vector. There is a problem, though: it so happens that, as one can easily verify, $\Phi_\mp = 0$, and consequently
\be
h_{\mp}^{\mu}{}_{\nu} = 0.
\ee
This means that the third-order solution does not contribute to the metric perturbation. As a matter of fact one can see that all odd-order terms in the metric perturbation vanish identically:
\be
h_{\mbox{\tiny{(5)}}}^{\mu}{}_{\nu} = h_{\mbox{\tiny{(7)}}}^{\mu}{}_{\nu} = 
h_{\mbox{\tiny{(9)}}}^{\mu}{}_{\nu} = \cdots = 0.
\ee

\subsection{Fourth-order gravitational waves}

At the fourth order, the sourceless gravitational field equation is
\begin{equation}
R_{\mq \mu \nu} = 0
\label{dH4}
\end{equation}
with $R_{\mq \mu \nu}$ the third-order Ricci tensor. In terms of the components of the metric perturbation, and using for $h_{\mt}^{\mu}{}_{\nu}$, $h_{\my}^{\mu}{}_{\nu}$ and $h_{\mp}^{\mu}{}_{\nu}$ the solutions already obtained, the field equation (\ref{dH4}) assumes the form
\be
\Box h_{\mq \mu \nu} -
\partial_\mu (\partial_\alpha h_\mq^{\,\alpha}{}_\nu - \onehalf \partial_\nu h_\mq ) -
\partial_\nu (\partial_\alpha h_\mq^{\,\alpha}{}_\mu - \onehalf \partial_\mu h_\mq ) = 
\textstyle \frac{1}{2} \, \Phi_\my^{\,2} k_\mu k_\nu \exp[{ i 4 k_\rho x^\rho}]
\label{dH4bis}
\ee
where $h_\mq = h_\mq^{\,\alpha}{}_\alpha$ and $\Phi_\my$ is given by Eq.~(\ref{fi2}).
On the other hand, at this order the harmonic coordinate condition (\ref{HarmoCoor}) is
\be
\eta^{\mu \nu} \, \Gamma_\mq^\lambda{}_{\mu \nu} +
\big(h_\my^{\mu \nu} + h_\mt^{\,\mu}{}_\alpha \, h_\mt^{\alpha \nu} \big) \, 
\Gamma_\my^\lambda{}_{\mu \nu} = 0.
\ee
Substituting the Christoffel connections, it becomes
\be
\partial_\mu h_\mq^{\,\mu}{}_\lambda - {\textstyle{\frac{1}{2}}} \partial_\lambda h_\mq =
{\textstyle{\frac{i}{8}}} \Phi^{\,2}_\my k_\lambda \exp[{ i 4 k_\rho x^\rho}].
\label{hc4}
\ee
In these coordinates, the fourth-order wave equation assumes the form
\begin{equation}
\Box\,  h_{\mq}^{\mu}{}_{\nu} = - \, {\textstyle{\frac{1}{2}}} \, \Phi^{\,2}_\my \, k^\mu k_\nu
\exp[{ i 4 k_\rho x^\rho}].
\label{ffee4}
\end{equation}
A monochromatic wave solution to this equation is
\be
h_{\mq}^{\, \mu}{}_{\nu} = \left(A_{\mq}^{\,\mu}{}_{\nu} + i B_{\mq}^{\,\mu}{}_{\nu} \right)
\exp[{ i 4 k_\rho x^\rho}]
\label{PhysGraWave4}
\ee
where
\be
A_{\mq}^{\,\mu}{}_{\nu} = {\textstyle{\frac{1}{16}}} \, \Phi^{\, 2}_\my \, \delta^\mu_\nu
\qquad \mbox{and} \qquad 
B_{\mq}^{\,\mu}{}_{\nu} = - \, {\textstyle{\frac{1}{32}}} \, \Phi^{\, 2}_\my \, 
\frac{K_\theta x^\theta}{K_\sigma k^\sigma} \, k^\mu k_\nu
\label{B4}
\ee
are the fourth-order amplitudes, with $K_\theta$ an arbitrary wave number four-vector.
The amplitude of the first part of the solution satisfies
\be
A_\mq^{\,\mu}{}_{\mu} \equiv A_\mq = \textstyle{\frac{1}{16}} \, \Phi^2_\my \qquad \mbox{and} \qquad
k_\mu A_{\mq}^{\,\mu}{}_{\nu} = \frac{1}{4} \, k_\nu A_\mq
\ee
whilst the amplitude of the second part of the solution satisfies
\be
B_\mq^{\,\mu}{}_{\mu} \equiv B_\mq = 0 \qquad \mbox{and} \qquad k_\mu B_{\mq}^{\,\mu}{}_{\nu} = 0.
\label{ttPart4}
\ee

We consider now a laboratory proper frame~---~endowed with a Cartesian coordinate system~---~from which the wave will be observed. If the wave is traveling in the $z$ direction of the Cartesian system, the wave vector is that given by Eq.~(\ref{kzdir}). In this case, the amplitude $A_{\mq}^{\mu}{}_{\nu}$ turns out to be
\begin{gather}
\left( A_{\mq}^{\,\mu}{}_{\nu} \right) = \frac{1}{64} \,  \Phi^{\,2}_\my \left(
\begin{matrix}
 1 & 0 & 0 & 0 \\
 0 &1 & 0 & 0 \\
 0 & 0 & 1 & 0 \\
 0 & 0 & 0 & 1
\end{matrix}
\right)
\end{gather}
from where we see that this part of the solution has just one physical component:
\be
A^{\;t}_{\mq t} = A^{\;x}_{\mq x} = A^{\;y}_{\mq y} = A^{\;z}_{\mq z} \equiv \textstyle{\frac{1}{64}} \, \Phi^{\,2}_\my.
\ee
On the other hand, choosing the arbitrary wave vector $K_\rho$ in such a way that $K_0 = K_1 = K_2 = 0$, the amplitude $B_{\mq}^{\,\mu}{}_{\nu}$ turns out to be
\begin{gather}
\left( B_{\mq}^{\,\mu}{}_{\nu} \right) = - \, \frac{1}{32} \frac{\Phi^{\,2}_\my z \, \omega}{c} \left(
\begin{matrix}
 1 & 0 & 0 & -1 \\
 0 & 0 & 0 & 0 \\
 0 & 0 & 0 & 0 \\
 1 & 0 & 0 & -1
\end{matrix}
\right).
\label{B4z}
\end{gather}
From this expression we see that this part of the solution has two physical components:
\be
B^{0}_{\mq 0} = - B^{z}_{\mq z} \qquad \mbox{and} \qquad B^{0}_{\mq z} = -
B_{\mq z}{}^{0}.
\label{PhysCom4}
\ee

\subsection{Higher-order gravitational waves}

Based on the solutions found up to here, one can infer that the generic form of a higher-order solution is  
\begin{equation}
h_{(2n)}^{\ \mu}{}_{\nu} = \Big(A_{(2n)}^{\ \mu}{}_{\nu} + i B_{(2n)}^{\ \mu}{}_{\nu} \Big)
\exp[{ i 2 n k_\rho x^\rho}]
\label{PhysGraWaven}
\end{equation}
where
\begin{gather}
\Big( A_{(2n)}^{\ \mu}{}_{\nu} \Big) = a_{2n} \Phi_\my^{n} \left(
\begin{matrix}
 1 & 0 & 0 & 0 \\
 0 &1 & 0 & 0 \\
 0 & 0 & 1 & 0 \\
 0 & 0 & 0 & 1
\end{matrix}
\right)
\end{gather}
and
\begin{gather}
\left( B_{(2n)}^{\ \mu}{}_{\nu} \right) = b_{2n} \, \frac{\Phi_\my^{n} z \, \omega}{ c} \left(
\begin{matrix}
 1 & 0 & 0 & -1 \\
 0 & 0 & 0 & 0 \\
 0 & 0 & 0 & 0 \\
 1 & 0 & 0 & -1
\end{matrix}
\right)
\label{B2zBis}
\end{gather}
with $a_{2n}$ and $b_{2n}$ numerical coefficients. Using Einstein's equations and the harmonic coordinate conditions, given respectively by
\begin{equation}
R_{(2n)}{}_{\mu\nu} = 0 \qquad \mbox{and} \qquad 
\Gamma_{(2n)}{}_{\nu} = 0,
\end{equation}
these coefficients can be easily computed. The first few of them are
\begin{eqnarray}
&&a_{6}=\frac{13}{4608} \qquad \qquad ~ b_{6} = -\frac{1}{1536} \nonumber\\
&&a_{8}=-\frac{83}{131072} \qquad ~~ b_{8} = -\frac{65}{147456}\\
&&a_{10}=\frac{5023}{31457280} \qquad b_{10} = \frac{163}{3145728} \nonumber \;.
\end{eqnarray}
And so on to any higher order. Like the second and fourth-order waves, all higher-order solutions have three independent physical components.

\section{Higher-order curvature invariants}

In general relativity, it is possible to construct two invariants in terms of the components of the Riemann curvature tensor.\footnote{In the general case there are actually three invariants, which degenerate in two for sourceless solutions to Einstein equation, which is our interest here.} They are the Kretschmann ($K$) and the Pontryagin ($P$) invariants, defined respectively by
\[
K = R_{\alpha \beta \mu \nu} R^{\alpha \beta \mu \nu} \qquad \mbox{and}
\qquad P = {R}_{\alpha \beta \mu \nu} {^{\star}R}^{\alpha \beta \mu \nu}
\]
with ${^{\star}R}^{\alpha \beta \mu \nu} = \onehalf \epsilon^{\mu \nu \rho \sigma}
R^{\alpha \beta}{}_{\rho \sigma}$ the curvature Hodge dual. The first is a scalar whereas the second is a pseudo-scalar quantity. Now, as is well-known, for linear gravitational waves, both Kretschmann and Pontryagin invariants are found to vanish identically~\cite{I1,I2,I3,I4}:
\[
K_\my \equiv R_\mt^\alpha{}_{\beta \mu \nu} R_{\mt\alpha}{}^{\beta \mu \nu} = 0
\qquad \mbox{and} \qquad 
P_\my \equiv {^{\star}R}_\mt^\alpha{}_{\beta \mu \nu} R_{\mt\alpha}{}^{\beta \mu \nu} = 0.
\]
The question then arises: is this a specific property of linear gravitational waves, or it is a general property of waves of any order? Using the higher-order solutions obtained in the previous section, we are able to give an answer to this question.

To begin with, using the first and second-order solutions, the third-order Kretschmann and Pontryaguin invariants are found to vanish identically,
\be
K_\mp \equiv 2 R^\alpha_{\mt \beta \mu \nu} R_{\my \alpha}{}^{\beta \mu \nu} = 0 
\ee
and
\be
\qquad P_\mp \equiv {}^\star R^\alpha_{\mt \beta \mu \nu} R_{\my \alpha}{}^{\beta \mu \nu} +
{}^\star R^\alpha_{\my \beta \mu \nu} R_{\mt \alpha}{}^{\beta \mu \nu} = 0.
\ee
Before considering the fourth-order invariant, it is important to remark that, although the third-order solution $h_{\mp\mu\nu}$ vanishes identically, the third-order Riemann tensor does not vanish because there are contributions to it coming from $h_{\mt\mu\nu}$ and $h_{\my\mu\nu}$. With this proviso, the fourth-order curvature invariants are then written as 
\be
K_\mq \equiv R^\alpha_{\my \beta \mu \nu} R_{\my \alpha}{}^{\beta \mu \nu} +
R^\alpha_{\mt \beta \mu \nu} R_{\mp \alpha}{}^{\beta \mu \nu}
\ee
and
\be
P_\mq \equiv {}^\star R^\alpha_{\my \beta \mu \nu} R_{\my \alpha}{}^{\beta \mu \nu} +{}^\star R^\alpha_{\mt \beta \mu \nu} R_{\mp \alpha}{}^{\beta \mu \nu} + R^\alpha_{\mt \beta \mu \nu} {}^\star R_{\mp \alpha}{}^{\beta \mu \nu}.
\ee
Performing these computations, both fourth-order curvature invariants are also found to vanish 
\begin{equation}
K_\mq=0 \qquad \mbox{and} \qquad  P_\mq=0.
\end{equation}
The above pattern repeats for any higher order. For example, as already discussed, the fifth-order metric perturbation is found to vanish: $h_{\mf\mu\nu} = 0$. Using this result it is then possible to compute the fifth-order curvature invariants, which are found to vanish,
\begin{equation}
K_\mf = 0 \qquad \mbox{and} \qquad  P_\mf = 0.
\end{equation}
Analogously, the sixth-order invariants can also be shown to vanish,
\begin{equation}
K_\ms = 0 \qquad \mbox{and} \qquad  P_\ms = 0.
\end{equation}
We can then conclude that both Krets\-chmann and Pontryagin invariants vanish for gravitational waves of any order.

\section{Discussion}

We have developed a systematic study of higher-order gravitational waves with the purpose of exploring their properties and the structure of the perturbation scheme. Considering that the first-order solution is computed in harmonic coordinates, and taking into account that such solution appears in all higher-order equations, the same class of coordinates must be used for dealing with higher-order gravitational waves. Furthermore, within this class, a specific coordinate system is then chosen, in terms of which the first-order perturbation $h_\mt^{\mu \nu}$ becomes transverse and traceless. In this case, as is well-known, the harmonic coordinate condition is equivalent to the divergence-free condition~(\ref{1oHC}). However, as we have explicitly verified, at orders higher than one the harmonic coordinate condition does not generalize to~(\ref{hoHC}), but to expressions with non-vanishing right-hand side. For example, at the second-order the harmonic coordinate condition assumes the form~(\ref{hc2}). The use of the correct harmonic coordinate condition is crucial for the mathematical consistency of the higher-order solutions.

An intriguing result of these higher-order solutions is that the second-order solution does not represent corrections to the first-order wave, but rather shows up as a completely different solution. In contrast, all higher than second-order solutions do represent corrections to the second-order wave. The first-order gravitational wave, therefore, stands apart from higher-order waves. This picture bears a strong resemblance to the case of nonlinear surface waves in shallow water, where solitary waves can be obtained from the Navier-Stokes equation (for an inviscid fluid) through a perturbation scheme \cite{dodd}. At first order one obtains a linear wave-equation whose solution does not represent a physical wave; its unique role is to determine the dispersion relation of the system. At the second order, one obtains a nonlinear evolution equation---the so-called Korteweg-de Vries equation---whose solution represents the solitary wave observed in nature. Like in the gravitational case, the second-order solution does not represent a correction to the first-order solution, but shows up as a completely different wave. Higher-orders terms of the perturbative scheme, however, do yield corrections to the second-order wave.

Another interesting feature of the higher-order perturbation analysis is that, except for the first order, all odd-order terms of the metric perturbation vanish identically. This feature may be related to the physical properties of the gravitational interaction. In fact, it is well-known that, for a spin-2 gauge boson, the particle-particle static interaction is always attractive in the linear approximation \cite{Kibble}. On the other hand, if presented in the perturbative series, odd-order terms would contribute with a repulsive static interaction between particle-antiparticle. That the perturbative series of Einstein equation naturally exclude all odd-order terms of the expansion can then be considered an evidence that the static interaction between matter-antimatter is always attractive at any order. It should be remarked that, in spite of compelling arguments favouring the idea that the static particle-antiparticle gravitational interaction should also be attractive, this is as yet an open question in the sense that it has never been verified experimentally. We mention in passing that there are currently some experiments aiming exactly at probing the gravitation behaviour of antimatter. One example is the ALPHA experiment at CERN \cite{alpha}.

A crucial output of the perturbation scheme is that, although the first-order solution is transverse and traceless, all higher-order solutions are not. In fact, the generic form of the higher-order metric perturbation is
\begin{equation}
h_{(2n)}^{\ \mu}{}_{\nu} = \Big(A_{(2n)}^{\ \mu}{}_{\nu} + i B_{(2n)}^{\ \mu}{}_{\nu} \Big)
\exp[i 2 n k_\rho x^\rho]
\end{equation}
with
\begin{equation}
A_{(2n)}^{\ \mu}{}_{\nu} = a_{2n} \Phi_{(2)}^{\,n} \, \delta^\mu_\nu
\qquad \mbox{and} \qquad
B_{(2n)}^{\ \mu}{}_{\nu} = b_{2n} \, \Phi_{(2)}^{\,n} \, 
\frac{K_\theta x^\theta}{K_\sigma k^\sigma} \, k^\mu k_\nu.
\end{equation}
We can thus clearly see that, while the $B$-part is transverse and traceless, the $A$-part is neither transverse nor traceless:
\begin{equation}
k_\mu \, A_{(2n)}^{\ \mu}{}_{\nu} = a_{2n} \Phi_{(2)}^{\,n} \, k_\nu
\qquad \mbox{and} \qquad
A_{(2n)}^{\ \mu}{}_{\mu} = 4 \, a_{2n} \Phi_{(2)}^{\,n}.
\end{equation}
As a consequence, the whole solution, that is, the solution that includes higher-order contributions, is neither transverse nor traceless. This result may eventually have important consequences for quantum gravity, and in particular for the definition of graviton itself. For example, in addition to the two degrees of freedom associated to massless fields, there is also a scalar degree of freedom associated to the non-vanishing trace of the solution.

On the other hand, when considering gravitational waves, since their sources are at enormous distance from Earth, the situation is somewhat different. To begin with, let us remark that, although plane waves have constant amplitude, the spherical symmetry of physical gravitational waves implies that their amplitudes fall off with the distance from the source. The amplitude of the first-order solution, for example, scales with distance $z$ according to $A_{\mt}^{\mu}{}_{\nu} \sim 1/z$. Considering that second-order effects are proportional to $A^{\,\rho}_{\mt \sigma} A^{\,\sigma}_{\mt \rho}$, it is usually argued that second-order gravitational waves fall off with distance as $1/z^2$, and are for this reason neglectful \cite{SecondOrder}. This is the case of the $A$-part of the second-order solution (\ref{PhysGraWave}), whose amplitude falls off as
\be
A_{\my}^{\,\mu}{}_{\nu} \sim 1/z^2.
\ee
However, due to an additional linear dependence on the distance, the amplitude of the $B$-part of that solution falls off as
\be
B_{\my}^{\,\mu}{}_{\nu} \sim 1/z. 
\ee
Contrary to the usual belief, therefore, second-order effects may be relevant even at large distances from the source. Neglecting the faster decaying $A$-part, the dominant solution at large distances is
\be
\tilde h_{\my}^{\,\mu}{}_{\nu} \simeq i B_{\my}^{\,\mu}{}_{\nu} \, e^{i 2 k_\rho x^\rho}.
\label{dom}
\ee
From the second-order solution~(\ref{ttPart}) we see that this wave is transverse and traceless. Its two physical components are
\be
\tilde h^{\,0}_{\my 0} = - \tilde h^{z}_{\my z} \qquad \mbox{and} \qquad \tilde h^{\,0}_{\my z} = - \tilde h_{\my z}{}^{0}.
\label{PhysCom}
\ee
For astrophysical and cosmological applications, therefore, a kind of screening mechanism eliminates the traceful part of the wave.

Finally, as a byproduct, we have studied the Kretschmann and Pontryaguin invariants. It is well-known that both invariants vanish identically for first-order gravitational waves. By using the higher-order solutions obtained in this paper, we have shown that these invariants vanish actually for gravitational waves of any order.

\section*{Acknowledgments}
The authors would like to thank FAPESP, CNPq and CAPES for financial support.


\end{document}